\documentclass[5p, authoryear]{elsarticle}

\usepackage[draft]{hyperref}
\usepackage{graphicx}
\usepackage{amsmath}
\journal{Astronomy \& Computing}

\newcommand{\specialcell}[2][c]{%
	\begin{tabular}[#1]{@{}c@{}}#2\end{tabular}}

\begin{document}

%%%%%%%%%%%%%%%%%%%%%%%%%%%%%%%%
%%%%%%%%%%%%%%%%%%%%%%%%%%%%%%%%
% Front Matter
%%%%%%%%%%%%%%%%%%%%%%%%%%%%%%%%
%%%%%%%%%%%%%%%%%%%%%%%%%%%%%%%%

 \begin{frontmatter}
 \title{AstroStat - A VO Tool for Statistical Analysis}
 \author[iuc]{Ajit K. Kembhavi}
 \ead{akk@iucaa.ernet.in}

 \author[cal]{Ashish A. Mahabal}
% \ead{aam@astro.caltech.edu}
 
 \author[iuc]{Tejas Kale}
 %\ead{kaletejas2006@gmail.com}
 
  \author[iuc]{Santosh Jagade}
%  \ead{santoshj@iucaa.ernet.in}
  
 \author[iuc]{Ajay Vibhute}
% \ead{ajay@iucaa.ernet.in}

\author[iuc]{Prerak Garg}

  \author[iuc]{Kaustubh Vaghmare\corref{cor1}}
 \ead{kaustubh@iucaa.ernet.in}
 
 \author[iuc]{Sharmad Navelkar}
 %\ead{sharmad@iucaa.ernet.in}
 
 \author[sym]{Tushar Agrawal\fnref{fn1}}
 
 \author[cu]{Asis Chattopadhyay}
 %\ead{akcstat@caluniv.ac.in}
 
 \author[jhu]{Deoyani Nandrekar\fnref{fn2}}
 
 \author[syn]{Mohasin Shaikh\fnref{fn2}}

 \cortext[cor1]{Corresponding Author}
 \fntext[fn1]{Author was affiliated with Persistent Systems Ltd. during the development of AstroStat.}
 \fntext[fn2]{Author was affiliated with the Inter-Unversity Centre for Astronomy and Astrophysics during the development of AstroStat.}
 \address[iuc]{Inter-University Centre for Astronomy and Astrophysics, Pune, India}
 \address[cal]{California Institute of Technology (Caltech), Pasadena, USA}
\address[sym]{Symantec, Pune, India}
\address[cu]{Department of Statistics, Calcutta University, Kolkota, India}

 \address[jhu]{John Hopkins University, Baltimore, Maryland, USA}
 \address[syn]{Synygy, Pune, India}
 
 \begin{abstract} 
AstroStat is an easy-to-use tool for
performing statistical analysis on data. It has been designed to be compatible
with Virtual Observatory (VO) standards thus enabling it to become an integral
part of the currently available collection of VO tools. A user can load data in
a variety of formats into AstroStat and perform various statistical tests using
a menu driven interface. Behind the scenes, all analysis is done using the public
domain
statistical software - R and the output returned is presented in a neatly
formatted form to the user. The analyses performable include exploratory tests,
visualizations, distribution fitting, correlation \& causation, hypothesis
testing, multivariate analysis and clustering. The tool is available in two
versions with identical interface and features - as a web service that can be
run using any standard browser and as an offline application.  AstroStat will provide an easy-to-use interface which can allow for both fetching data and performing power statistical analysis on them.

\end{abstract}

 \end{frontmatter}
 
%%%%%%%%%%%%%%%%%%%%%%%%%%%%%%%%%%
%%%%%%%%%%%%%%%%%%%%%%%%%%%%%%%%%%

%%%%%%%%%%%%%%%%%%%%%%%%%%%%%%%%%%
% Actual Document Begins from Here.
%%%%%%%%%%%%%%%%%%%%%%%%%%%%%%%%%%

%%%%%%%%%%%%%%%%%%%%%%%%%%%%%%%%%%
% INTRODUCTION
%%%%%%%%%%%%%%%%%%%%%%%%%%%%%%%%%%

\section{Introduction} 
AstroStat\footnote{http://voi.iucaa.ernet.in:8080/astrostat} is a powerful VO
compatible tool, developed by the Virtual
Observatory-India (VOI) project, for statistical analysis of data.  It provides
a number of statistical  tests, ranging from the simple to the more complex and
sophisticated, which are performed using a very simple to use graphical
interface.   The analysis is carried out using the highly developed statistical
package R, which is available in the public domain.   AstroStat uses in-built
graphics for easy visualisation of the data as well as the results of the tests
performed.   It incorporates  various VO  standards, so that it can easily be
linked to a wide range of VO tools like the plotting and visualisation tools
VOPlot and TOPCAT and can use the Astronomical Data Query Language to obtain
data from VO compatible services for statistical analysis.

AstroStat has evolved from the statistical analysis tool VOStat, which was first
developed through a collaboration between groups from Caltech and Pennsylvania
State University and  later through collaboration between these two groups and
VOI. VOStat is available as a web-service from the Centre for Astrostatistics at
Penn State\footnote{http://astrostatistics.psu.edu:8080/vostat/}.   AstroStat
has been developed as an independent tool by VOI, in collaboration with a group
from Caltech, with important inputs from various astronomers, statisticians and
software engineers. 

The AstroStat code is made of two parts - the main backbone code written in Java and the R snippets which are made available to the user when a test is run. Both these codes are being made available to the community under GNU GPL license agreement.\footnote{The source code can be obtained by mailing a request to voindia@iucaa.ernet.in}

The present article is organized as follows. In
Section 2, we provide an overview of the tool and in Section 3, the details of R
as a statistical backend are discussed. In Section 4 and 5, we cover the inner
implementation details of AstroStat including descriptions of various VO
standards. In Section 6 we provide an illustrative application of AstroStat
and in Section 7 briefly discuss future directions. 

%%%%%%%%%%%%%%%%%%%%%%%%%%%%%%%%%%

%%%%%%%%%%%%%%%%%%%%%%%%%%%%%%%%%%
% Overview of AstroStat
%%%%%%%%%%%%%%%%%%%%%%%%%%%%%%%%%%
\section{An Overview of AstroStat}

AstroStat comes in two flavors - an offline version\footnote{IMPORTANT: The AstroStat stand-alone or offline version is still in development. While the application can still be downloaded from http://voi.iucaa.ernet.in/$\sim$voi/AstroStat.html, it is not yet ready for the end user.} bundled in the form of an
executable Java Archive (.jar) and a web version which can be run in any standard
browser. The interface, which has been designed with ease-of-use in mind, has
been kept the same in both the versions. The primary interface comprises of three
ever-present sections - i) which enables the user to load data, ii) a collection
of tests categorized into Exploratory, Advanced and Expert, and iii) a help
section which presents a description of the currently selected test with
examples and any extra notes. A fourth section appears on selecting a test and
this provides options to select and transform columns, supply necessary
parameters to the test (eg. type of correlation when computing a correlation
matrix), choose the nature of output etc. \ref{fig:screenshot}. 

\begin{figure}[h!] 
\begin{center}
\includegraphics[width=0.5 \columnwidth]{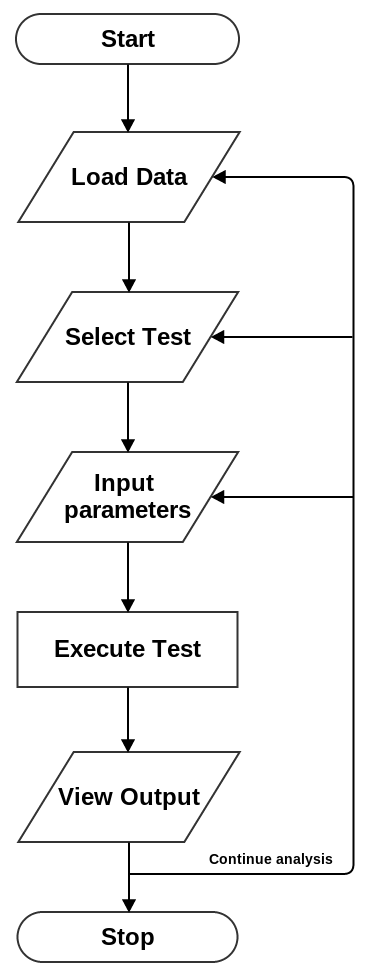}
\caption{A flow chart illustrating the user perspective of the workflow in
AstroStat.} 
\label{fig:userflow}
\end{center} 
\end{figure}

The typical workflow, from the end user's perspective, 
is shown in Figure \ref{fig:userflow}. The user first loads data into the
application, in the form of a file either on the local hard
drive or on a web server. Data can also be loaded using the Table Access
Protocol (TAP) \citep{tap} or through Simple Access Message Protocol (SAMP) \citep{samp}, as described in detail
in Section 5. It is
possible to load more than one file at a time and a list of all loaded files is available in the form a drop-down menu. As a next step, the
user selects one of the three categories of tests and a test within it. A
complete list of all tests available can be found in Appendix A. The Help
section updates itself to reflect the currently selected test and offers a quick
overview of what the test does, possible examples and special notes, if any.
When a test is selected, the fourth section appears where a user
inputs parameters required by the test. Once done, the user clicks \emph{Run
Test} and AstroStat performs the analysis and displays the output in a tabular
form with tooltips to aid interpretation.

Since all the four sections described above are always visible, the user can
easily run another test or the same test with modified input parameters, or
refer to
the help section for a quick reminder of say, what exactly the output means,
etc. The output is in a friendly and neatly formatted form and can be easily
saved. The plots can be saved into a single ZIP file while the tables and other
output data can be stored in a plain ASCII text format.

In addition to these features, AstroStat also offers other
functional features like

\begin{itemize} 
\item A quick-look summary statistics pop-up for the currently
loaded data.  
\item Ability to view both the tabular version and the original
file. This allows the user to ensure that the data have been loaded correctly.
\item The user can define new columns by performing common operations on
existing columns. (e.g. sum of two columns, square of a column, etc.) 
\item One
click access to the VOPlot service \citep{voplot} for interactive plotting and
data
visualization.  
\item Ability to view the R code used in the actual analysis so
that a user may build upon this code for further work. If the user wishes to modify the R code provided to perform further analysis, this will have to be done outside of AstroStat in a R shell. The R code is provided under the GNU GPL license. At the time of writing this article is being written, the R code provided by the web version to the user includes a lot of code which is especially needed for a seamless interaction between AstroStat and R. In a future release, we will clean the code being served to the user so it can become easier for the user to modify it.
\end{itemize}

In the subsequent sectons, we describe the detailed implementation and features
of the tool.

%%%%%%%%%%%%%%%%%%%%%%%%%%%%%%%%%%

%%%%%%%%%%%%%%%%%%%%%%%%%%%%%%%%%%
% Statistical Backend
%%%%%%%%%%%%%%%%%%%%%%%%%%%%%%%%%%

\section{Statistical Backend}

The R language \citep{ihaka96} came into existence as a free
counterpart of the S statistical language from Bell Labs. Like S, R \citep{R} has all the
common tools needed for advanced statistics: linear and non-linear modeling,
various statistical tests, time series analysis, classification, clustering etc.
Ross Ihaka and Robert Gentleman developed R with user participation in mind
which has resulted in a very large number of contributions from the users. The
Comprehensive R Archive Network (CRAN)\footnote{http://cran.r-project.org/}
hosts the
user packages and has easy interfaces to download and install any of the
packages from geographically distributed mirror sites. In early 2014 the count
has crossed 5000 packages. As it is arguably the most versatile open-source
system for statistics we decided to use it as the backend for the
AstroStat service. The original collaboration for developing such a
service was between Caltech and Penn State with the coding to be done at Caltech
\citep{mahabal2002, graham2005}. Part of the undertaking was to provide users
with a
set of tools as well as broad and basic guidance about which tool to use under
specific conditions. This is important given that newer packages keep entering
CRAN everyday and it can be bewildering for new users to choose from competing
packages. 

We have categorized the functionality of AstroStat into exploratory,
advanced, and expert. We provide an overview of the tests in this section, while
greater detail is provided in the Appendix. The exploratory set contains
descriptive statistics features such as plotting histograms of single variables,
making simple x-y plots of one parameter against another, pairs' plot to obtain
x-y
plots for several variables, box-plots, and obtaining basic statistics such as
mean and standard deviation. Analysis of Variance (ANOVA) and sample generation
are also included. The
advanced set contains line- and plane-fitting through simple- and multiple
linear regression analyis; correlation matrix, covariance analysis,
Kolmogrov-Smirnov test (both one- and two-sample) etc. The expert
set allows multivariate classification with Hierarchical clustering, K-means
partitioning and clustering, kernel smoothing, as well as tasks that can help
with censored data like survival analysis. The help files about the
tests have text and links explaining when specific tests can be used.

One of the difficulties in using R is that the syntax often does not parallel
that of other
languages that users typically encounter. By providing a Graphical User Interface (GUI) we remove the need
for the user to start coding in R. At the same time we provide the R code that
generated the analysis so that the user can learn from there. If the user is
already well-versed with R, then this code will allow her to further analyse
similar data independently.  Another difficulty is that the same functionality
in
different types of figures uses different keywords, making the learning curve
steeper. By providing plots at the click of a button, we ensure that users do
not have to wrestle with those differences. In addition, instead of using the
base graphics which have the above problems, we have adopted the \emph{ggplot2}
\citep{ggplot2} library which is more uniform. 

The ggplot2 library by Hadley Wickham\footnote{http://ggplot2.org} is based on
the
Grammar of Graphics \citep{wilkinson2005}. This is a layered approach to
graphics
which allows the user to trivially add and subtract different layers to the
plot. For example, if one wants to plot data from some part of the sky with
different magnitude ranges (e.g. from synoptic surveys such as Digital Access to
a Sky Century \@ Harvard - DASCH - at the bright end, intermediate Catalina
Real-Time Transient Survey - CRTS - at the intermediate range, and simulated
Large Synoptic Survey Telescope - LSST - at the deep end), one can make separate
layers for the three sets. The error-bars and fits/contours  can be additional
layers, and any subset of these can be plotted. Since these exist as layers,
another subset can be equally easily plotted without having to go through the
entire process of reading files, assigning data etc. For any data set, this is
achieved by defining \textit{mappings} from \textit{data} to aesthetic
attributes of geometric objects, \textit{geoms}, like points and lines. These
can then be included in statistical transformations (\textit{stats}) in specific
coordinate systems. Further, \textit{faceting} (aka conditioning) allows easy
subsetting. Finally \textit{scale} and \textit{coord} allow it to be rendered on
to a plot exactly the way a user wants to.  While the powerful statistical
techniques of R are used in the analysis, it is the versatile ggplot2 that
provides
visualization that is crucial, especially in the initial aspects of a project
when the workflow is still being crystalized. We also provide ggplot2 code when
plots are generated, allowing the users to learn advanced plotting through R on
the go as well. On account of appearance and associated aesthetics alone,
ggplot2
is superior, but programmers who would want to build further on the layered
approach will thus find it very rewarding. As a bonus, defaut figures generated by {\it ggplot2} are near-publication quaility and just a small number of tweaks make them fully so. Going in to those details is beyond the scope of this article but can be found at several places on the internet. 

ggplot2 makes beautiful but static plots. As a result dynamically changing axis names, labels etc. is not possible. In the Appendix B, we provide a comparison of plots and the code needed to generate them using the default graphics library of R and the ggplot2 library being used by AstroStat. While ggplot2 does have a layered approach to graphics, we'd like to note that the AstroStat user does not have a direct access to these layers. The R code will, however, allow the user some level of access to the R users.

Finally, a note on the scalability issues concerning R. As bulk of the analysis in AstroStat is done by R, R largely determines the scalability of the application. It is not possible to quote a stringent limit on how big a data can be loaded in R. This will be a function of the resources available on the machine on which AstroStat is being run. Subject to community response, we can look into the possibility of making AstroStat compatible with variants of R specifically designed for use with large data in parallel computation environments.

%%%%%%%%%%%%%%%%%%%%%%%%%%%%%%%%%

%%%%%%%%%%%%%%%%%%%%%%%%%%%%%%%%%
% Implementation Details
%%%%%%%%%%%%%%%%%%%%%%%%%%%%%%%%%

\section{Implementation Details} 
\subsection{Input} 

AstroStat accepts data in
three file formats: VOTable, ASCII, and FITS (binary). As mentioned earlier,
files can be loaded in two ways, either from the local hard drive or from a web
server by specifying the URL. Data in VOTable (discussed in Section 5) and FITS
formats are loaded automatically since these store detailed metadata in an
unambigious way. However, when loading ASCII files (.csv, .tsv, etc.), a data
parser module is invoked which requests certain inputs from the user to enable
accurate loading of the data. The queries posed to the user are - 

\begin{itemize} 
\item Are column names, their data types, units, and/or
UCDs specified in the file? If yes, what are their respective line numbers?
\item From which line does the actual data begin?  
\item Which character should
be interpreted as a comment character?  
\item What is the delimiter which
separates individual entries in a single row?  
\item Has the tool correctly
identified the data type of every column? If not, the user may specify the
correct types.  
\end{itemize}
 
In most cases, the data parser module will be able to find the details on its
own and thus this step is more about the user confirming automatically
discovered parameters than supplying actual information. After loading a file,
the data can be viewed in a tabular format by clicking on the `View Data' option
in the toolbar. In the Data Input panel, a selection of statistics of every
column, including minimum, mean, median, variance, and maximum, can be quickly
viewed by hovering the mouse pointer over `Data Summary'. 

There are two additional input methods available. The user can use an existing
VO compliant tool that supports the Simple Access Messaging Protocol (SAMP),
which enables control and data communication between two user applications, to
load data directly into AstroStat. Or a user may click the ``TAP" button on the
toolbar and use the Table Access Protocol (TAP) which allows the user to use the
Astronomical Data Query Language (ADQL) \citep{adql} to query data from a
compatible data
service and make it available directly to the application. A detailed discussion
on input via SAMP and TAP is deferred to Section 5.

As the next step, the user selects a test category and then a test and this
refreshes the Input Panel to display all relevant and possible inputs. As
mentioned before, the tests have been catalogorized into Exploratory, Advanced
and Expert. For every test, necessary and relevant inputs are sought to tailor
the analysis to the user's demands. At the same time, all these inputs have
default values for quick analysis. The inputs for every test have been
thoughtfully curated to maximize the flexibility as well as convenience for the
user.

\begin{figure*}[h!] 
\begin{center}
\includegraphics[width=15cm, keepaspectratio=true]{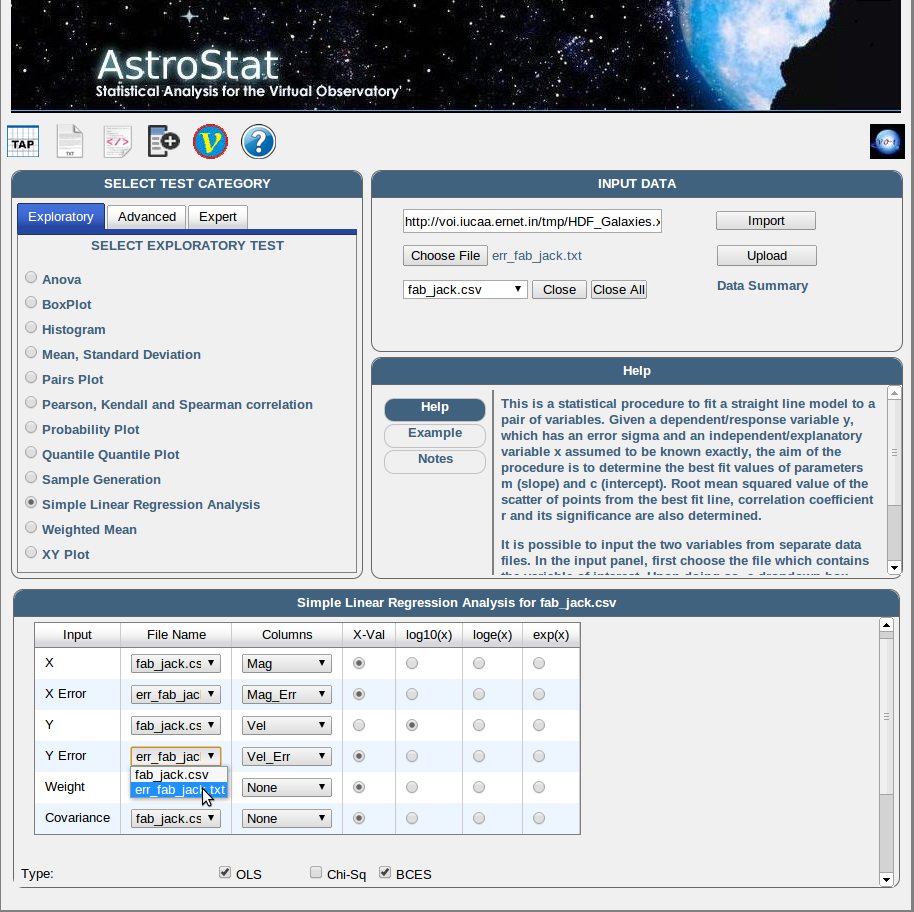}
\caption{A screenshot of the web version of AstroStat showing the toolbar at
the top and the four sections that comprise the primary interface. It also illustrates a feature which allows a user to select columns from multiple files.} 
\label{fig:screenshot}
\end{center} 
\end{figure*}

	To illustrate the features of the Input Panel, Figure
\ref{fig:screenshot} shows a snapshot of the panel for \emph{Simple Linear
Regression}. The user is first prompted to select columns which will act as $Y$
(dependent variable), $X$ (independent variable), $Y_{\rm{error}}$,
$X_{\rm{error}}$, etc in the analysis. Transformation of some or all variables
is possible by clicking on the appropriate radio button(s) adjacent to the
column names. Finally, choices are sought for the type(s) of regression analysis
to be performed, the format of output plots, and whether the user desires to
obtain bootstrap error estimates. 

On clicking \emph{Run Test}, the generated output is displayed as a new tab (in
case of the Web version) or a new window (in case of the offline version), so
that all input parameters remain available for the user to cross-check or rerun
the test after tweaking the parameters. The main application window also has an
option to add a new column to the loaded table. On selecting this option, the
user is prompted with a dialog box using which a new column can be created by
combining the existing columns in any arbitrary mathematical expression. Such a
feature can be useful, for example, when computing residuals for the derived
best-fit line for further analysis.

\subsection{Output}

	The output provided by any test in R, the statistical backend in
AstroStat, can appear cluttered and non-intuitive for a user unaccustomed to the
nuances of the language. Hence, under the hood, AstroStat performs intensive
processing and reformatting of this output to display the most relevant bits of
information. In general, the following tenets are followed when displaying the
output:

\begin{itemize} 

\item Display output in a tabular format for ease of
understanding and clarity.  
\item Separate the output window into two sections:
one for displaying the textual output and the other for showing plots associated
with the analysis.  
\item Distinctly specify important input information like
data variables selected for analysis, sample size, function evaluated, etc.
\item Wherever applicable, provide supplementary information in a tabular format
for further analysis. For example, on performing principal component analysis,
the PCA scores are available for download in ASCII format for visualization in
VOPlot or any other plotting tool. 		
\item Every output table has a
\emph{?} symbol associated with it which reveals a tooltip that gives a quick
explanation of the parameters listed.  

\end{itemize}
	
The output window also comes with a toolbar which offers the following features.
\begin{itemize} 
\item \emph{R Code}: View/Download the code used to perform the
analysis. The code can be used to subsequently perform more complex analysis
using R or as an aid for learning R.  
\item \emph{Save}: Save tab-separated,
tabular output in an ASCII file.  
\item \emph{Plots}: Save all plots (if any) in
a ZIP file.  
\item \emph{Table}: Save output table (if any) in a
comma-separated, ASCII file.  
\item \emph{VOPlot}: Send data used in the
analysis to VOPlot for (further) visualization.  
\end{itemize}

\subsection{Inner Workings}

\begin{figure*}[h!] 
\begin{center}
\includegraphics[width=15cm]{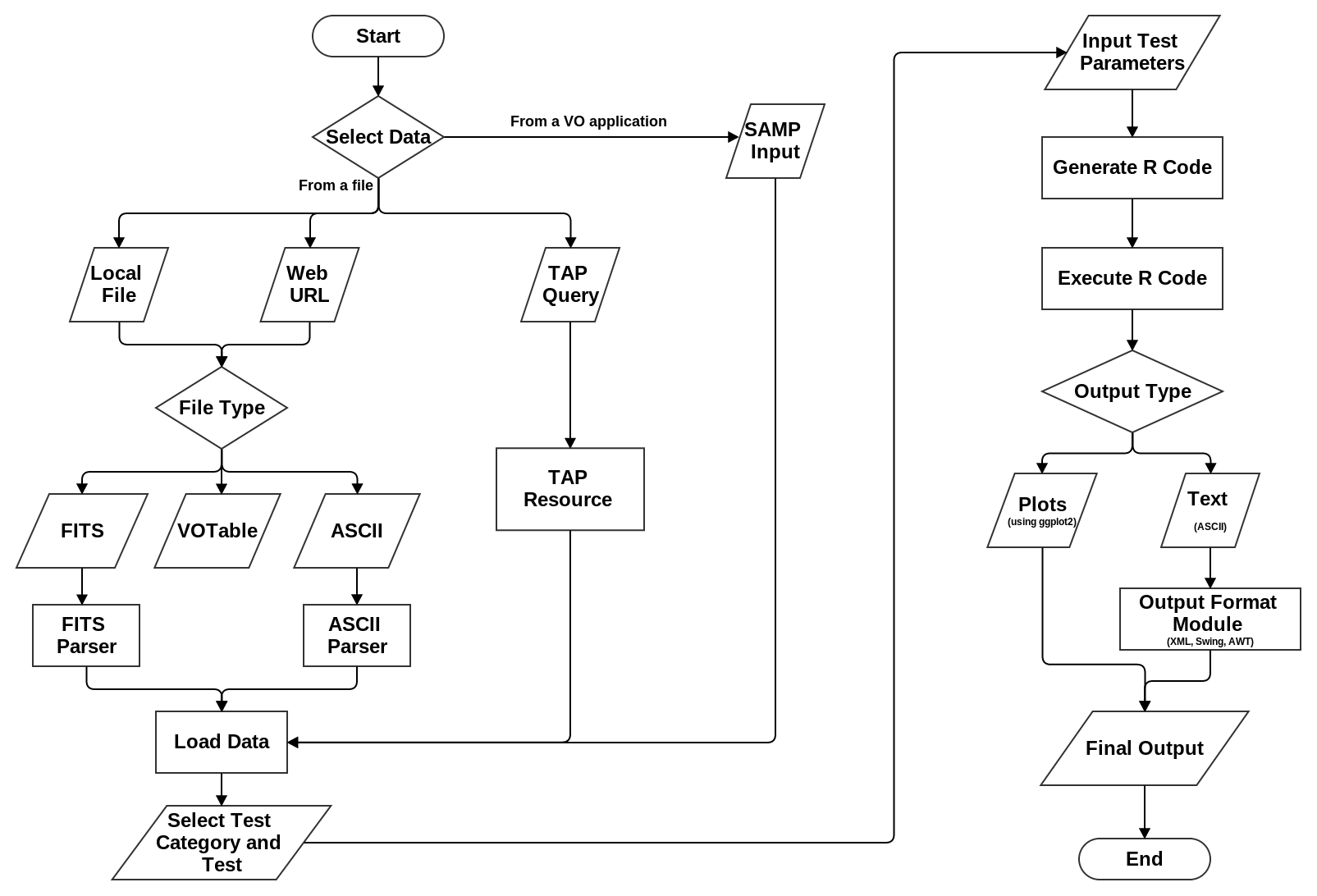}
\caption{A flowchart showing the inner workings of AstroStat.} 
\end{center} 
\label{fig:programmerflow}
\end{figure*}

From a user's perspective, the workflow described in Figure \ref{fig:userflow}
is sufficient. For someone wanting to understand the details of the inner
workings of AstroStat, Figure \ref{fig:programmerflow} gives a clearer
picture. The workflow is valid for both the web and the offline versions. A few
details of platforms and technologies used are described below.

The web version was largely programmed using Java Server Pages (JSP). Like PHP,
it allows creation of rich dynamic web pages but uses the Java programming
language. A large collection of tag libraries allows clear separation
between the model and controller parts of the code. As the controller part is
implemented in Java, there is a large number of robust libraries and frameworks
available which can be easily plugged in or adopted. Further, there is
support for multi-threading, concurrency and background processing. Although the
current implementation does not take advantage of these features they could well
be used if such performance demands are expected from the service by the
community.

While JSP has been used for the overall user interface design, the web version
also makes use of the Yahoo User Interface (YUI) libraries which enable a clean
and highly appealing display of the output in tabular form. The validation of
all information entered by the user is done using code written in Javascript.
The web server hosting AstroStat is located at IUCAA in Pune, India. Any
information entered by the user is transmitted to the Virtual Observatory India
(VOI) server which runs a Servlet that generates an appropriate R script. The
execution is then carried out by a Java system call and any output produced is
converted to XML format and sent to the user. A Javascript then parses this XML
output to generate the final formatted output which is displayed.

The stand-alone version is largely based on Java's Abstract Window Toolkit
(AWT). The AWT is a part of the Java Foundation classes and is frequently used
to design GUIs. The stand-alone version works in a
similar manner as that of the web version sans the client-server communication
mechanism. This version requires R to be installed on the local machine and any
output generated by the R script is directly formatted into the final form. The
output tables are also implemented using AWT. Keeping up with the spirit that a user
does not have to know R in order to use AstroStat, the application is able to
locate the R installation except in some cases where a user may be required to
provide the R installation path. Further, some of the tests use extra packages
available on CRAN. Again, the tool allows the user to download any missing
packages from CRAN directly. All dependencies can be installed in one go using
an option from the menu or on-demand, whenever a user tries to run a test that
requires a particular package.

%%%%%%%%%%%%%%%%%%%%%%%%%%%%%%%%%

%%%%%%%%%%%%%%%%%%%%%%%%%%%%%%%%%
% Implementation of VO Standards
%%%%%%%%%%%%%%%%%%%%%%%%%%%%%%%%%

\section{Implementation of VO Standards in AstroStat}

Being able to share astronomical data seamlessly across a variety of data
services and analysis tools is at the heart of the Virtual Observatory. The data
can be in the form of images, spectra and/or tables. Thus the International
Virtual Observatory Alliance (IVOA) has explored and adopted various standards
over the years to enable easy information sharing. The adoption of these
standards ensures that every data service in the world can "talk" to any other
such service effortlessly and share information. This allows individual
developers to create specialized tools that can perform specific types of
analyses. Since all tools support these standards, these tools together should
be able to serve most needs of an astronomer. The primary motivation in creating
AstroStat has been to provide a service capable of taking data from any VO
compatible data service or source and perform various statistical tests on them.
Thus some essential standards have been implemented in both the versions of
AstroStat.

\subsection{VOTable} 
This is an XML based standard created for storage of
tabular data. A VOTable \citep{votable} can be viewed as an unordered collection
of rows, with
the description of each row contained in the metadata. Each row can be viewed as
a collection of cells, each containing one element of a primitive data type. The
VOTable was designed to be a very flexible format with astronomical tables in
mind. As it is XML based, one can take advantage of Extensible Stylesheet
Language Transformations (XLST) which allows for easy transformation of data from one
form to another.

The design philosophy of VOTable has been motivated by large data use cases and
distributed computing in mind. For example, it is possible for a VOTable to
contain only metadata with a link to the actual data stored on a web server. The
data part in turn can be in pure XML format (called TABLEDATA) generally used in
the case of a small number of rows, and FITS binary format. The metadata is
allowed to
be semantically rich through the use of standards such as Uniform Content
Descriptor (UCD) \citep{ucd}, Utype, Units and Space Time Coordinate (STC)
\citep{stc}. 

AstroStat accepts VOTable input as a preferred or default input type and further
includes parser modules for processing FITS files and ASCII tables which are in
common use among astronomers.

\subsection{SAMP} 
SAMP \citep{samp} stands for Simple Application Messaging Protocal. It was
developed as a standard way of allowing software tools to exchange both
control and data with each other. For example, one can imagine that while using
a tool such as VOPlot for visualizing data, some points of interest are noted in
the plot. It should be possible to select these points and enable another
completely different software application to, say, query and display specific
images of the corresponding astronomical object. SAMP, which is not specific to
the domain of VO or astronomy, provides a valuable binding layer between
user-centric applications. It is therefore possible for several independent
applications serving very specific purposes to work as an integrated whole.

AstroStat supports the SAMP protocol and thus any compatible tool can exchange
data with it. An in-built option in AstroStat loads the VOPlot service
and uses SAMP to have the current active data file loaded, thus allowing its use
for any kind of data visualization supported by VOPlot. 

\subsection{TAP} 
Table Access Protocol \citep{tap} allows astronomers to acquire tabular
data by writing queries as is done for data access from the Sloan Digital Sky
Survey (SDSS) or the UKIRT Infrared Deep Sky Survey (UKIDSS). The queries can be
written in Astronomy Data Query Language (ADQL) \citep{adql} which is a
standardized version
of the commonly used SQL. With AstroStat supporting TAP, it should be
straightforward to query a rich database supporting the TAP protocol from within
the application. The query will return a table which can be used in AstroStat
directly for analysis.

The option to use TAP can be invoked by clicking on ``TAP" tool button. The user
may either select an existing TAP compatible data service or search for one
based on keyword(s) or specify the URL of the service if available. Once a
compatible data server is selected, the user then selects a table and the
description of the metadata is presented. The metadata aids in the construction
of the desired ADQL query which, when submitted, returns a VOTable that either
can be saved locally or loaded into AstroStat. A few commonly used queries are
available in a dropdown list which can be used as starting points for building a
custom query.

TAP allows data querying and analysis to be integrated within a single
application. The intermediate steps of downloading, reloading and necessary
formatting of data are eliminated, thus making the workflow very fluid and
simple.

%%%%%%%%%%%%%%%%%%%%%%%%%%%%%%%%%

%%%%%%%%%%%%%%%%%%%%%%%%%%%%%%%%%
% CASE STUDIES
%%%%%%%%%%%%%%%%%%%%%%%%%%%%%%%%%

\section{Fundamental Plane - A Use Case}

In this section, we demonstrate a use case for AstroStat to study the
fundamental plane of elliptical galaxies, an important relation often discussed
in extragalactic astronomy. All calculations and plots in this section are made
using AstroStat.

The fundamental plane \citep{george87, dressler87} is a 3-dimensional linear
relation,
valid for elliptical galaxies and bulges of later type galaxies, which can be
written as 

\begin{equation} 
\log\ (\rm{r}_e) = A \left<\mu_e\right> +\ B\log\ \sigma_c + C
\end{equation}

\noindent where $\rm{r}_e$ is the bulge effective radius, $\left<\mu_e\right>$ is the average
surface brightness internal to $r_e$ and $\sigma_c$ is the central velocity
dispersion. The fundamental plane is important in practical terms because it
provides a technique for estimating distance to galaxies independent of their
redshift and theoretically because it provides insights into the dynamics of
galaxies.

\begin{figure}[h!] 
\begin{center} 
\includegraphics[width=0.9\columnwidth]{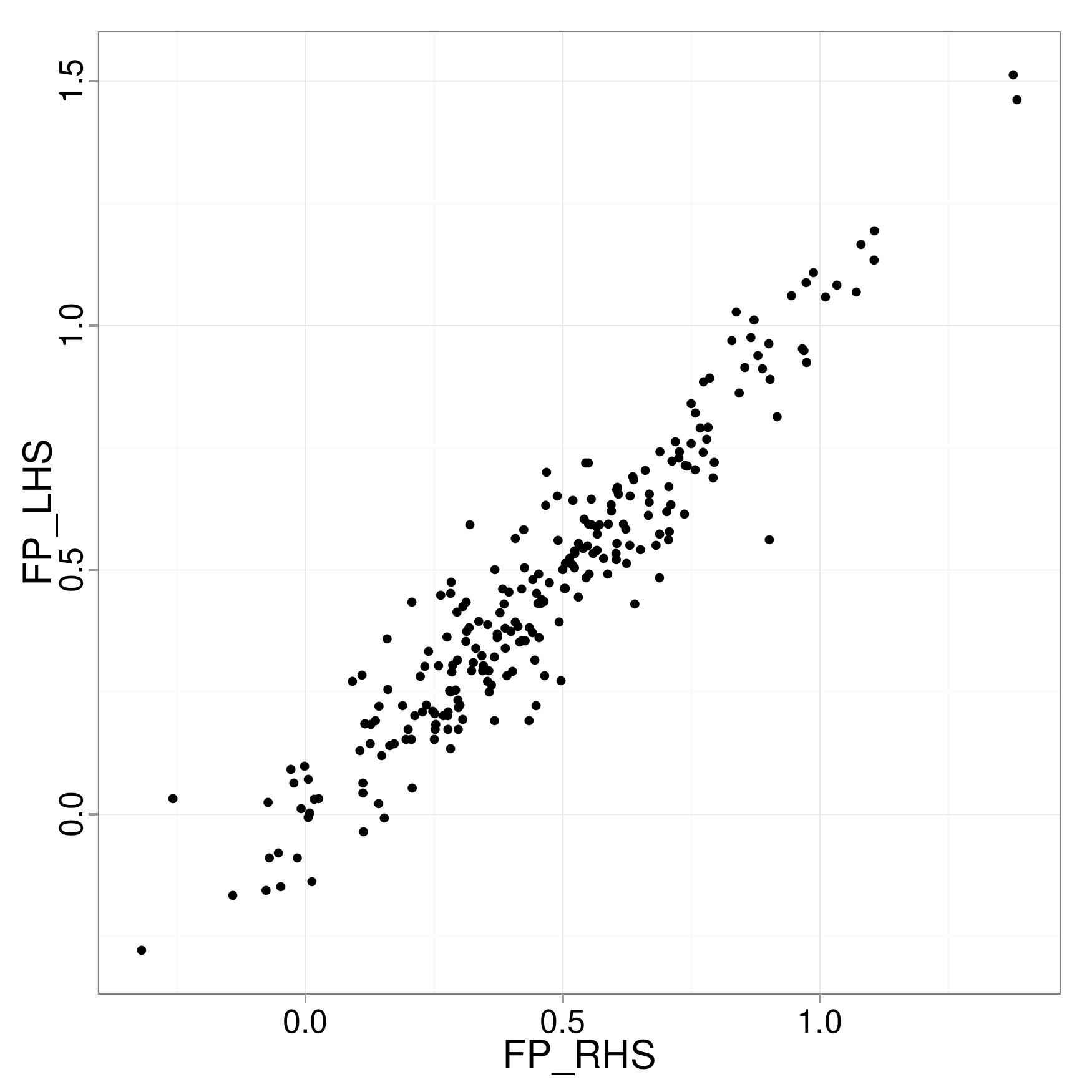} 
\caption{A plot between the left-hand-side vs the right-hand-side of the
Equation \ref{eqn:fp}, often referred to as the edge-on view of the plane.}
\label{fig:plane} 
\end{center} 
\end{figure}

\begin{figure}[h!] 
\begin{center} 
\includegraphics[width=0.9\columnwidth, height=07cm]{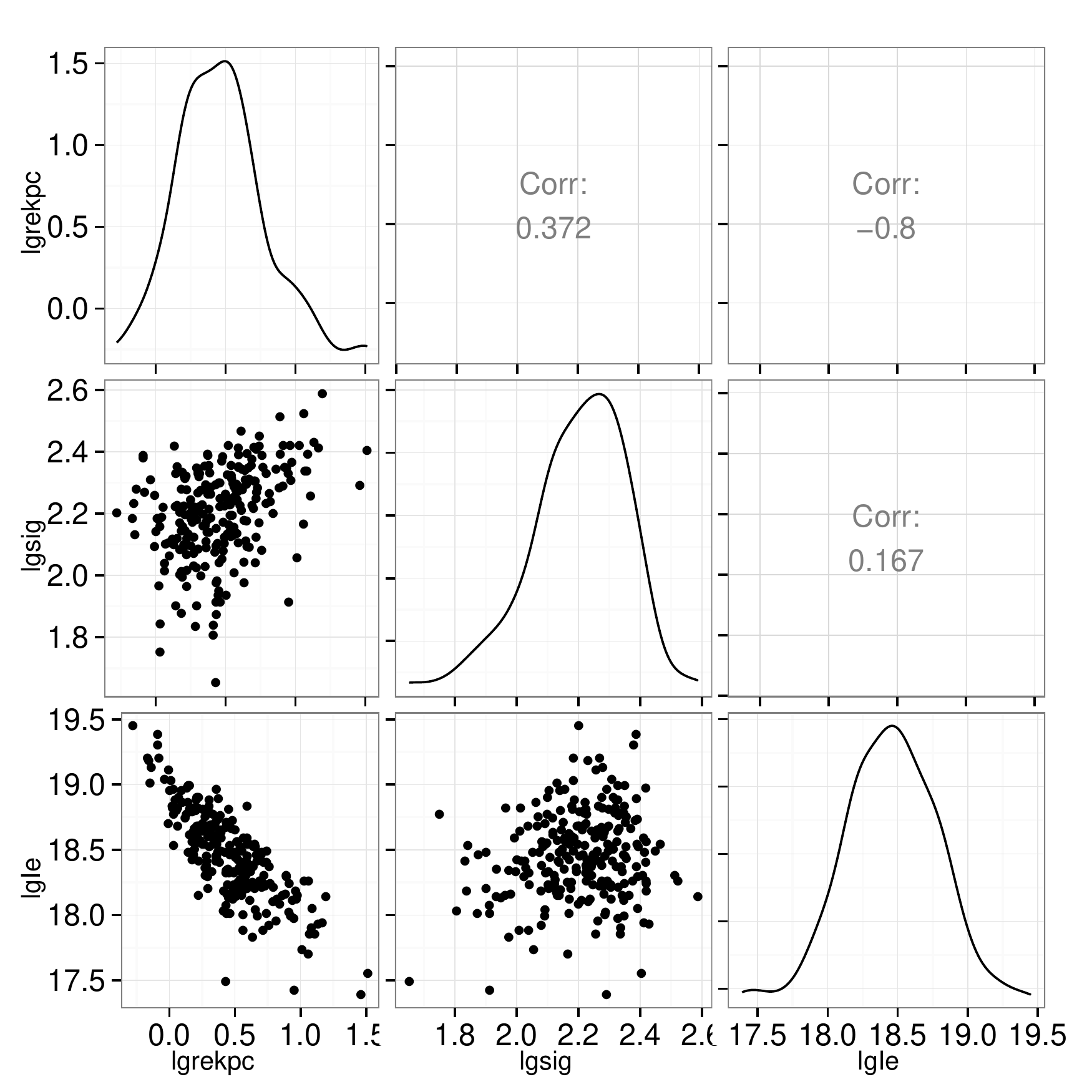} 
\caption{A pairs plot generated by AstroStat. The plot on the bottom-left
corner is the Kormendy plot which relates effective radius of the galaxy with
its average surface brightness.}
\label{fig:pairs} 
\end{center} 
\end{figure}

\begin{figure}[h!] 
\begin{center} 
\includegraphics[width=0.9\columnwidth]{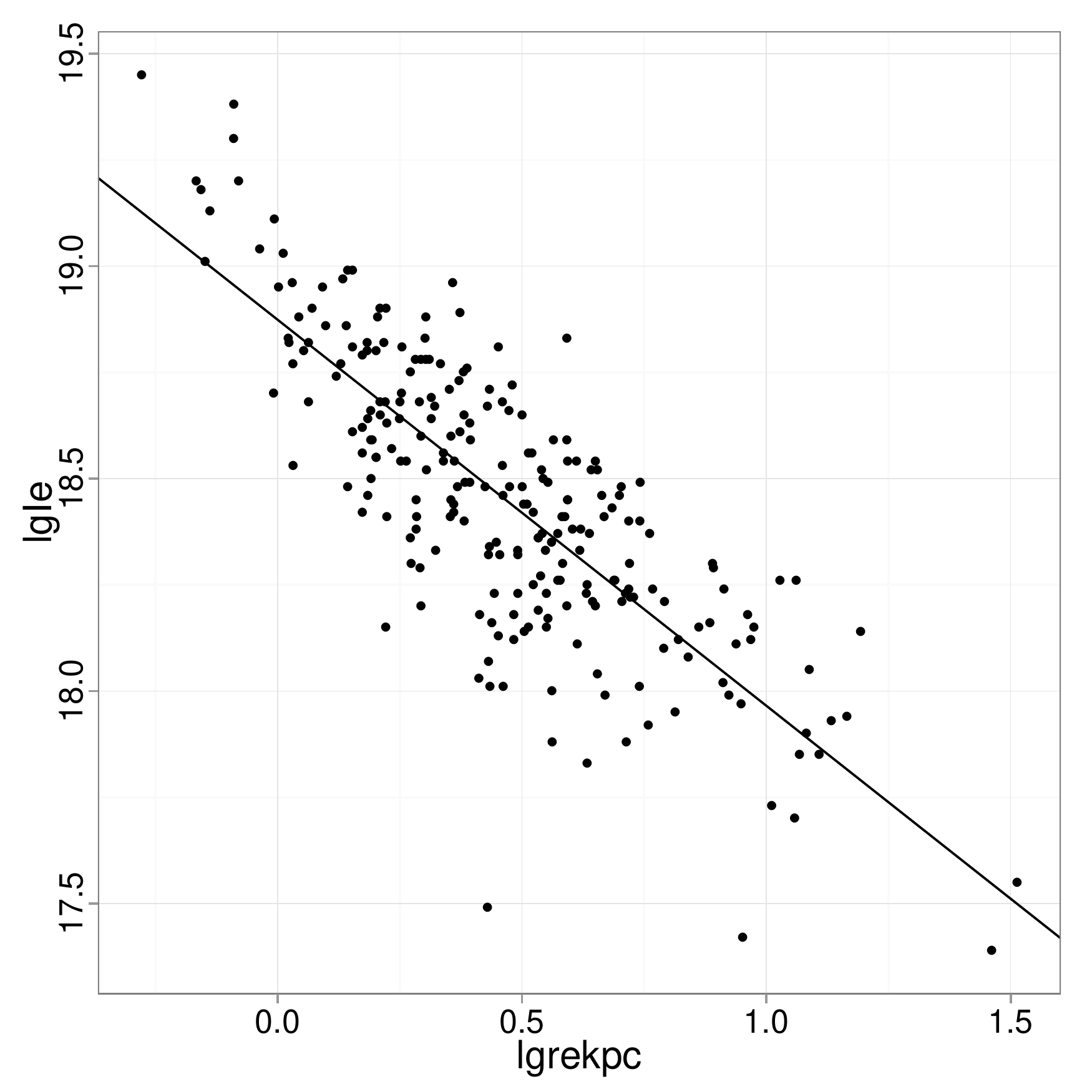} 
\caption{A plot showing the best-fit line between effective radius of the
galaxy and its average surface brightness i.e. the Kormendy plot.}
\label{fig:linefit} 
\end{center} 
\end{figure}

%\begin{figure*}[h!] 
%\begin{center} 
%\includegraphics[width=15cm]{reg_out.pdf} 
%\caption{A screenshot showing output of simple linear regression in AstroStat.}
%\label{fig:slr_result} 
%\end{center} 
%\end{figure*}

For the present illustration, we use a well known data set from \citet{jor96}
for 244 galaxies containing morphological parameters derived from images taken in the 
Gunn r-band. \footnote{If the reader wishes to perform all the steps on his/her
own, detailed instructions can be found at
http://voi.iucaa.ernet.in:8080/exercises/astrostat/fundamentalplane/}. The data set contains three columns viz. $r_e$, $\log I_e$ and $\sigma_c$. Here, $\log I_e$ is the log of the mean intensity within effective radius, which is same as $\left<\mu_e\right>$ within a scaling factor. Given such a data set, it is easy to determine the fundamental plane by performing multiple linear regression (under \emph{Advanced Tests}). The equation thus obtained is

\begin{equation} 
\log\rm{r}_e = 12.569 + 1.042 \log\ \sigma_c  - 0.780 \log \rm{I}_e
\label{eqn:fp}
\end{equation}

\noindent The original equation obtained by \citet{jor96} is as follows.

\begin{equation} 
\log\rm{r}_e = \rm{const.} + 1.240 (\pm 0.07) \log\ \sigma_c  - 0.82 (\pm 0.02)
\log \rm{I}_e
\label{eqn:jor}
\end{equation}

\noindent The differences in the coefficients arise due to the different approaches used to determine the best-fit coefficients. \citet{jor96} minimize the deviations along the orthogonal direction to the plane while AstroStat's (and R's) multiple linear regression routine minimize the deviation along the direction of the dependent variable ($\log r_e$, in this case). We have verified, using an independent program, that the result for \citet{jor96} can be exactly reproduced if the minimization is carried out along the orthogonal direction.

The edge-on view of the fundamental plane can be plotted as a simple XY plot between the left and right hand side of the equation \ref{eqn:fp}. The column creation feature can be used for creating a new column which represents the right hand side of the equation and XY plot option can be used to generate the final plot. This is shown in Figure \ref{fig:plane}. 

We have illustrated the determination of fundamental plane but this approach assumes prior knowledge of its existence. We now illustrate the process by which such a relation can be discovered \emph{ab initio} from the data.

To follow such a process, we start by making a \emph{pairs' plot}. This is a grid of plots which shows XY plots for different pairs of the variables present in the data. Since it is superfluous to plot a variable against itself, the plots along the diagonal are instead histograms of the variables. The plots in the upper diagonal half, to avoid reptition are filled with Pearson correlation coefficients for the correlation between each pair of variables. The Pearson correlation coefficient is a measure of the extent to which two variables are linearly correlated. The pairs' plot for the current data is shown in Figure \ref{fig:pairs}. Both visually as well as numerically, it can be seen that the correlation between $\log I_e$ and $\log r_e$ is the \emph{strongest} with a Pearson correlation coefficient of -0.8. The probability that such a correlation can arise by chance can be computed by determining the correlation matrix under \emph{Advanced Tests}. The output from this test also gives a matrix of p-values and for the pair of variables comprising the effective radius and the mean intensity within it, it is almost zero.  

\begin{table*}
\begin{center}
\begin{tabular}{|c|c|c|c|c|c|c|c|}
	\hline
	$X$ & $Y$ & Intercept & Slope & RMS Scatter & $r$ & $t$ & $p(>t)$ \\
	\hline
	$\left<\mu_e\right>$ & $\log \rm{r}_e$ & \specialcell{18.874\\ ($\pm0.024$)} & \specialcell{-0.9084\\ ($\pm0.044$)} & 0.205 & -0.800 & -20.722 & $< 2.2 \times 10^{-16}$ \\
%	& & ($\pm0.024$) & ($\pm0.044$) & & & & \\
	\hline
\end{tabular}
\label{table:linefit}
\caption{A table showing the output of Simple Linear Regression test in AstroStat. Here, $r$ refers to the Pearson's correlation coefficient, $t$ refers to the coefficient's test statistic, and $p(>t)$ refers to the p-value of the test statistic.}
\end{center}
\end{table*}

\begin{figure}[h!] 
\begin{center} 
\includegraphics[width=0.9\columnwidth]{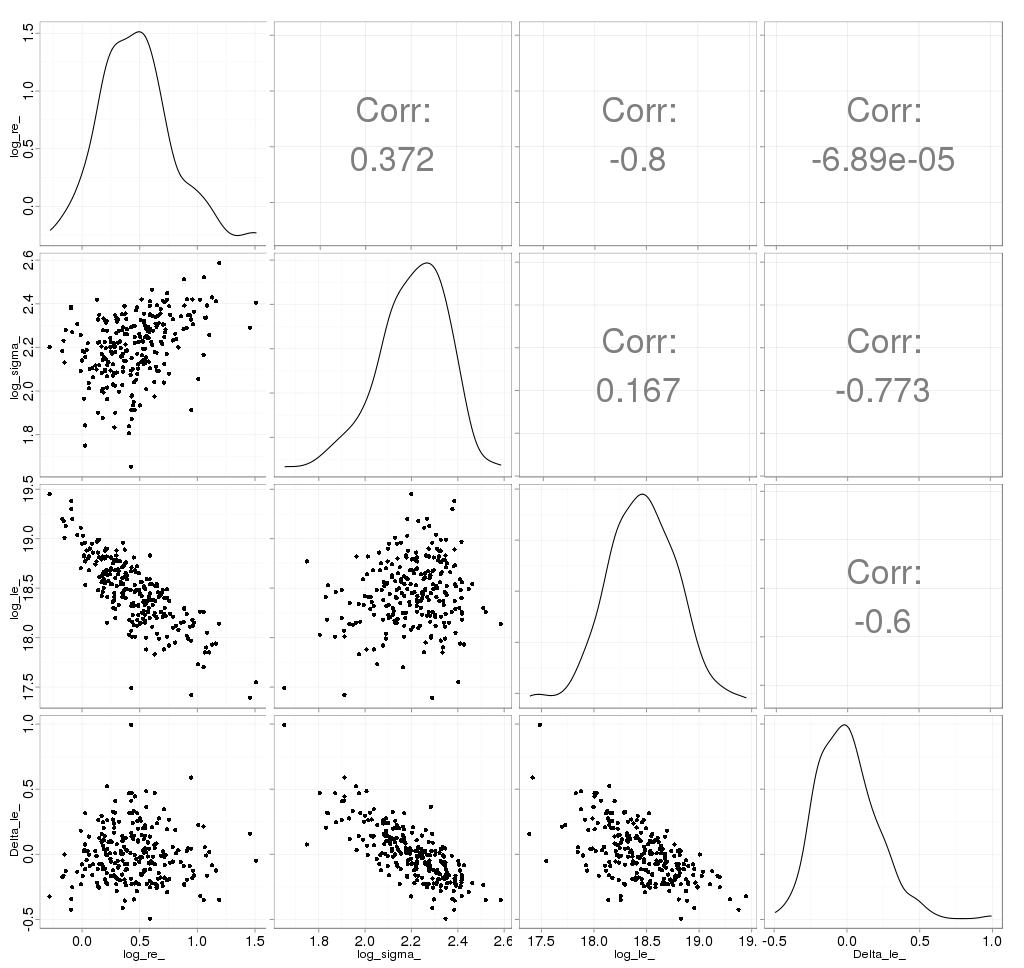} 
\caption{A plot between the left-hand-side vs the right-hand-side of the
Equation \ref{eqn:fp}, often referred to as the edge-on view of the plane.}
\label{fig:4plot} 
\end{center} 
\end{figure}

One can, at this point, fit a straight line to these two quantities. This can be done using \emph{Simple Linear Regression} test. The results for this test are summarized in Table 1 and the best-fit line is shown in Figure \ref{fig:linefit}. This is the well known Kormendy relation \citep{Kormendy77}. The root-mean-square (RMS) scatter in this correlation is 0.2. Can this scatter be explained using measurement errors? If the data also comprised of the error information, answering this question could be straight-foward but since no such information is available, we can use another approach to check whether the scatter is truly random. For this, we will define the deviation of the points from the best-fit line as ($y_i - a - bx_i$) and add this as a new column to our file. Once again, we make a Pairs' plot which results in a $4 \times 4$ grid of plots as shown in Figure \ref{fig:4plot}. This plot reveals a strong correlation between the deviations and $log \sigma_c$ with correlation coefficient of -0.773 and a p-value of 0 (as checked using the correlation matrix). This implies that the scatter in the Kormendy relation is not random but systematically arises from a third variable. This hints at a higher dimensional relationship which can be fitted using multiple linear regression as already shown above.

Another approach to arrive at the Fundamental plane of elliptical galaxies is to use Principal Component Analysis (PCA). PCA is a very powerful tool that can be used to study the relationships between the variables. A common use of PCA is to reduce the overall dimensionality of the data set by constructing new synthetic variables. The PCA test can be run from under \emph{Advanced Tests} in AstroStat. The output comprises two pieces of information - the component loadings and the total variance accounted by each component. The principal components obtained are in order of decreasing variance. 

The first principal component determined accounts for more than 80\% of the variance and is shown below.

\begin{equation}
 \rm{PC1} = -0.651 \log(\rm{r}_e) -0.027 \log\ \sigma_c  + 0.759 \log I_e
 \label{eqn:pc1}
 \end{equation}
 
\noindent The component loadings (coefficient of each term) indicate a strong correlation between $\log r_e$ and $\log I_e$ which is consistent with the above analysis. Now, the third principal component is the direction of minimal variance. If three quantities lie on a plane, the normal to the plane is the direction of minimal variance. Therefore, PC3 can be interpreted as a normal to the plane. We can further assume that the variance in the direction of PC3 is due to noise and thus the equation of plane can be written as

 \begin{equation}
 0.563 \log(\rm{r}_e) -0.687 \log\ \sigma_c  + 0.459 \log I_e = \rm{constant}
 \label{eqn:pc3}
 \end{equation}

\noindent Rearranging the terms in Equation \ref{eqn:pc3}, we get,

 \begin{equation} 
 \log(\rm{r}_e) =  + 1.220 \log\ \sigma_c  - 0.815 \left<\mu_e\right> +  \rm{constant}
 \label{eqn:pc3_2}
 \end{equation}

\noindent As can be seen, this fundamental plane relation reasonably agrees with
Equation \ref{eqn:fp}. It can
also be seen that it agrees \emph{more} with the original equation derived by
\citet{jor96}. This is because the Principal Component Analysis, by
construction, will minimize variance in an orthogonal direction. 

This example illustrates how a data set can be loaded in AstroStat and
subjected to various statistical tests allowing a user to gain insights about
the underlying correlations. That this data set need not sit on the user's
desktop but can be directly queried off Vizier or other services using the Table
Access Protocol tool makes it easy for astronomers to perform data querying and
analysis without leaving the web browser window. If the user wants to
dig deeper into the several options actually provided by R, or say, customize
the plots, the R code made avaiable can be used as a starting point.

%%%%%%%%%%%%%%%%%%%%%%%%%%%%%%%%%
% FUTURE WORK
%%%%%%%%%%%%%%%%%%%%%%%%%%%%%%%%%

\section{Future Work} 

The development of AstroStat has been made as modular as possible to allow for easy 
extensibility of the application's functions. If the
community of users requires inclusion of other commonly used analyses, for e.g.
those applicable to time series data, they can be easily added as additional
tests, perhaps even in a new test category, by the VOI development team. VOI is open to community feedback to drive the growth of AstroStat. Some evident future directions include modifying AstroStat to be compatible with the 'big data wave' and to present R code to the user in a fashion that it can be easily run and tweaked by the user on a local instance of R. A possible consideration for the future is to provide an interface by which the end-users will be able to add new R modules to AstroStat. 

At the time of writing this article, support for the web SAMP module is being tested. This will enable data from tools such as TOPCAT or Vizier to directly transmit tabular data to AstroStat or transmit tables loaded in AstroStat to other VO compatible tools. This can, for example, overcome the limitation of static plots provided by \emph{ggplot2} by allowing users to link data with a tool that supports advanced plotting such as VOPlot or TOPCAT. The reader is encouraged to watch out for new developments as well as offer suggestions for further development of the tool.

The current web version was not designed with touch-based interface in mind and
may be inconvenient to use on such devices. This has motivated us to create a lightweight
version of the AstroStat web application with a touch-friendly interface that
can work effortlessly on devices with limited computing resources. The
development of such a service is currently being planned.

Finally, an Android app is also being developed to provide 
a pedagogical interface to basic statistical analysis that can aid in
classroom teaching. The app will allow students to understand descriptive
statistics, correlation, straight-line
fitting, effects of outliers and visualize data using scatter plots, line
graphs and bar charts. Particular attention is being paid to make the app fully 
compatible with \emph{Aakash} tablets which are low-cost devices being
widely distributed in schools and colleges in India by the Ministry of Human
Resource Development, of the Government of India. 
%%%%%%%%%%%%%%%%%%%%%%%%%%%%%%%%%

%%%%%%%%%%%%%%%%%%%%%%%%%%%%%%%%%
% ACKNOWLEDGEMENTS
%%%%%%%%%%%%%%%%%%%%%%%%%%%%%%%%%

\section*{ACKNOWLEDGEMENTS}
The authors would like to thank Kiran Jappanwar for help with the development of
the original TAP module; Somak Raichaudhary \& Peter Tino, whose ideas helped
shape the interface of the tool; and Sajeeth Philip \& Sudhanshu Barway who
helped test the tool on a large scale and provided valuable feedback.
Well before IUCAA took over the current development, the initial VOStat work was done at Caltech. AAM acknowledges the support of National Science Foundation (NSF)
grant DMS-0101360. SG Djorgovski and MJ Graham also contributed to the initial design.
The VO-India project is a collaboration between the Inter-University Centre for
Astronomy and Astrophysics and Persistent Systems Ltd., under which
AstroStat has been developed, is partially funded by
the Ministry of Communications and Information Technology of the Government of
India. Kaustubh acknowledges financial assistance provided by the Council of
Scientific and Industrial Research (CSIR), India.

\section*{References}
\bibliographystyle{model2-names}
\bibliography{mybib}
%%%%%%%%%%%%%%%%%%%%%%%%%%%%%%%%%

%%%%%%%%%%%%%%%%%%%%%%%%%%%%%%%%%
% APPENDIX A
%%%%%%%%%%%%%%%%%%%%%%%%%%%%%%%%%

\section*{Appendix A} 

The list of statistical techniques available in AstroStat
is presented below: \footnote{Location of every test in the application is
highlighted using an alphabet next to it with the following key:
\emph{E}=Exploratory, \emph{A}=Advanced, \emph{X}=Expert} \begin{itemize} \item

\textbf{Exploratory Analysis}
		
	\begin{itemize} \item \emph{Boxplot (\emph{E})}: A graphical
representation of summary statistics for a variable.  \item \emph{Histogram
(\emph{E})}: Display the distribution of a variable as a histogram.  \item
\emph{Mean, Standard Deviation (\emph{E})}: A tabular depiction of mean and
various measures of variability of a variable along with its histogram and
boxplot.  \item \emph{Pairs Plot (\emph{E})}: A matrix of scatter plots for
selected variables.  \item \emph{Weighted Mean(\emph{E})}: Provide a customized
mean of N data points, each of which is scaled according to a given criterion.
\item \emph{XY Plot (\emph{E})}: A scatter plot of two variables.  \end{itemize}

	\item \textbf{Correlation and Causation}

	\begin{itemize} \item \emph{Pearson, Kendall, and Spearman Correlation
(\emph{E})}: Non-parametric methods to test the degree of correlation between
two variables.  \item \emph{Correlation Matrix (\emph{A})}: Provide correlation
between variables along with their significance.  \item \emph{Covariance
Analysis (\emph{A})}: Provide covariances between variables.  \item \emph{Simple
Linear Regression Analysis (\emph{E})}: Fit a straight line model to two
variables and determine the degree of correlation.  \item \emph{Multiple Linear
Regression Analysis (\emph{A})}: Fit a $n$-dimensional plane to $n$ variables
and determine the degree of correlation.  \item \emph{ANOVA (\emph{E})}: Compare
the means of two or more groups of a variable created using a specified
criterion.

	\end{itemize}

	\item \textbf{Fitting distributions}

	\begin{itemize} \item \emph{Probability Plot (\emph{E})}: A graphical
technique to determine whether a variable follows one of the provided
distributions.  \item \emph{Quantile-Quantile Plot (\emph{E})}: A graphical
technique to check whether the distributions of two variables are equivalent.
\item \emph{Empirical Distribution Function (\emph{A})}: Graphically depict the
estimate of an underlying cumulative distribution function obtained from a
sample.  \item \emph{Kernel Smoothing (\emph{X})}: Fit simple, localized models
to small subsets of observational data to obtain an estimate of the distribution
of a variable.  \end{itemize}

	\item \textbf{Hypothesis Testing}
	
	\begin{itemize} \item \emph{One and Two Sample t-Test (\emph{A})}: In
the One-sample case, estimate the probability of population mean being equal to
a specified value. In the Two-sample case, estimate the probability of
population means of two samples being equal.

\item \emph{Kolmogrov Smirnov One Sample Test (\emph{A}): A non-parametric test to determine if a variable follows a given distribution.}

\item \emph{Kolmogrov Smirnov Two Sample Test (\emph{A}): A non-parametric test to determine if two variables follows the same distribution.}

\item \emph{Testing for mean when
variance is known (\emph{A})}: A parametric method to test whether the mean of a
population is equal to a specified value when the population variance is known.
\item \emph{Wilcoxon Rank-Sum Test (\emph{A})}: A non-parametric test to
determine whether two sample distributions come from the same parent continuous
distribution.  \item \emph{Kruskal Wallis k-sample test (\emph{X})}: A
non-parametric comparison of the medians of two or more groups of a variable
created using a specified criterion.  \item \emph{Shapiro-Wilks Test for
Normality (\emph{X})}: A test to determine if a sample is drawn from a Gaussian
distribution.  \end{itemize}

	\item \textbf{Multivariate Analysis}

	\begin{itemize} \item \emph{Factor Analysis (\emph{A})}: A
dimensionality-reduction technique to identify the latent variables influencing
the data.  \item \emph{Independent Component Analysis (\emph{A})}: A
dimensionality-reduction technique for non-Gaussian data that extracts
statistically independent components (signals) from the data (source).  \item
\emph{Principal Component Analysis (\emph{A})}: A dimensionality-reduction
technique that finds linear combinations of variables that capture most of the
variation in the data.  \end{itemize}

	\item \textbf{Clustering}

	\begin{itemize} \item \emph{H-Clustering (\emph{X}}: Distribute data
points among a specified number of clusters using an appropriate dissimilarity
criterion.  \item \emph{k-Means Partitioning (\emph{X})}: Cluster data points
into a given number of clusters by optimizing their Euclidean distance from
cluster centers.  \item \emph{Optimum k for k-Means (\emph{X})}: Determine the
optimum number of clusters to be obtained using $k$-means clustering.
\end{itemize}

	\item \textbf{Miscelleneous}

	\begin{itemize} \item \emph{Sample Generation (\emph{E})}: Generate a
sample from one of the available distributions.  \item \emph{Survival Analysis
(\emph{X})}: Obtain survival curves for selected variables.  \end{itemize}
\end{itemize}

\section*{Appendix B}

In this section, a comparison is offered between plots and the code needed to produce these plots, for the standard base graphics library or R and the ggplot2 library, being employed by AstroStat. 

For this example, we use a sample data set from \emph{Modern Statistical Methods for Astronomy} by Feigelson \& Babu, described in section 6.9.1 pg139. A code is written to produce a plot which comprises two subfigures - a histogram  and quantitle plot of redshifts. The first part of the code generates the same plot using standard base graphics and the second using ggplot2. The code is sufficiently commented to illustrate the advantages of the ggplot2 code over its counterpart for the base graphics library.

First, we show the code needed to generate these plots using the base graphics library. 

\begin{verbatim}
## Generate histograms and QQ plots 
## using base and ggplot2 graphics

# Obtain a large sample of 
# SDSS quasar redshifts
qso <-  read.table(
"http://astrostatistics.psu.edu
/MSMA/datasets/SDSS_QSO.dat", head=TRUE)
z_all <- qso$z



# Plot a histogram and quantile plot of this 
# sample using base graphics 
par(mfrow=c(1,2)) # Set layout of plot window 

# Create histogram
hist(z_all, breaks="scott", main="", 
xlab="Redshift", col="black") 

# Create quantile plot
plot(
quantile(z_all, seq(1,100,1)/100, na.rm=TRUE)
,pch=20, cex=0.5, xlab="Percentile",
ylab="Redshift") 


# Reset plot window layout to default
par(mfrow=c(1,1)) 

\end{verbatim}

\begin{figure*}[h!] 
\begin{center} 
\includegraphics[width=0.9\textwidth]{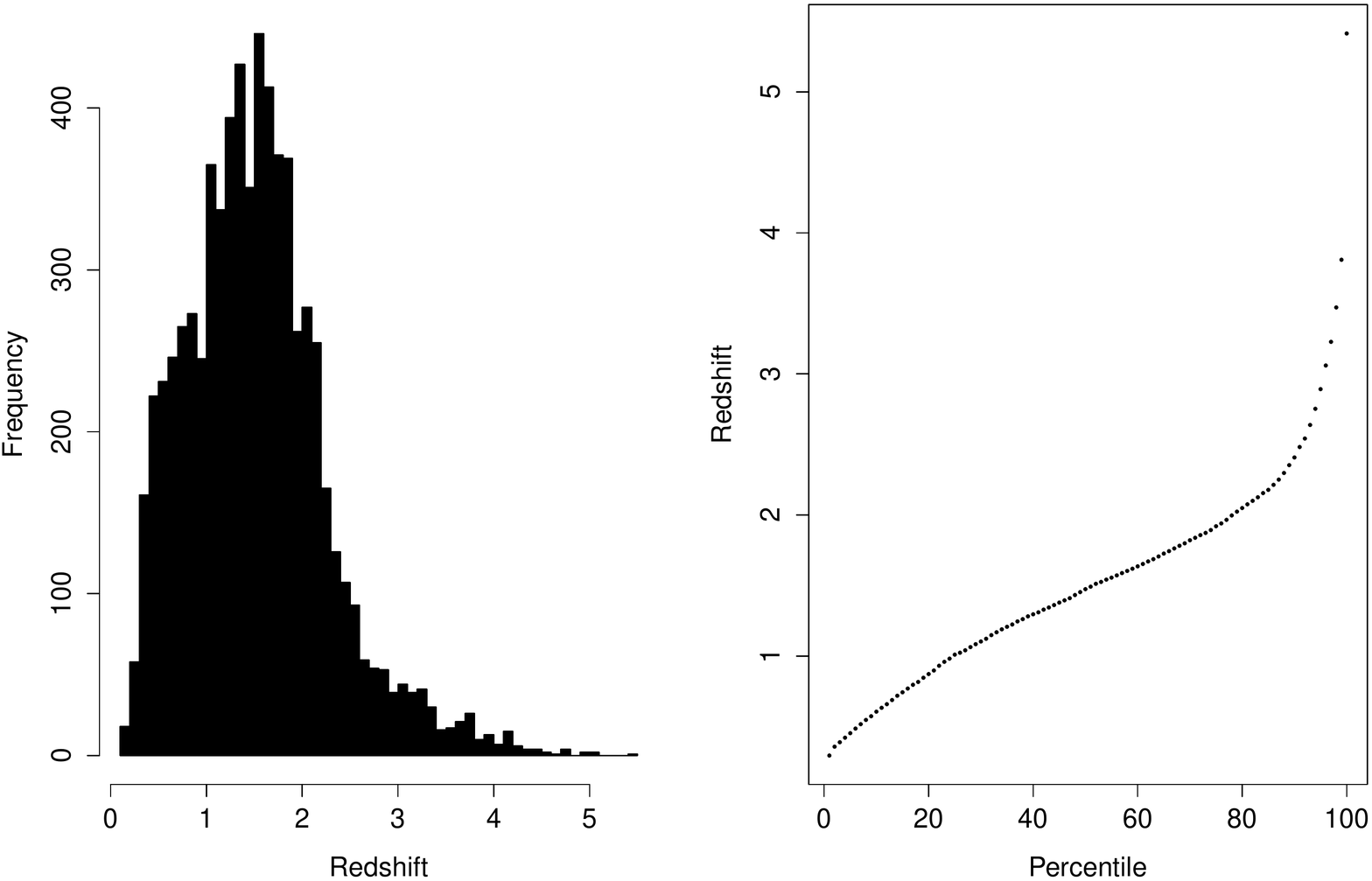} 
\caption{A histogram and a quantile plot generated using base graphics in R.}
\label{fig:base} 
\end{center} 
\end{figure*}

\begin{figure*}[h!] 
\begin{center} 
\includegraphics[width=0.9\textwidth]{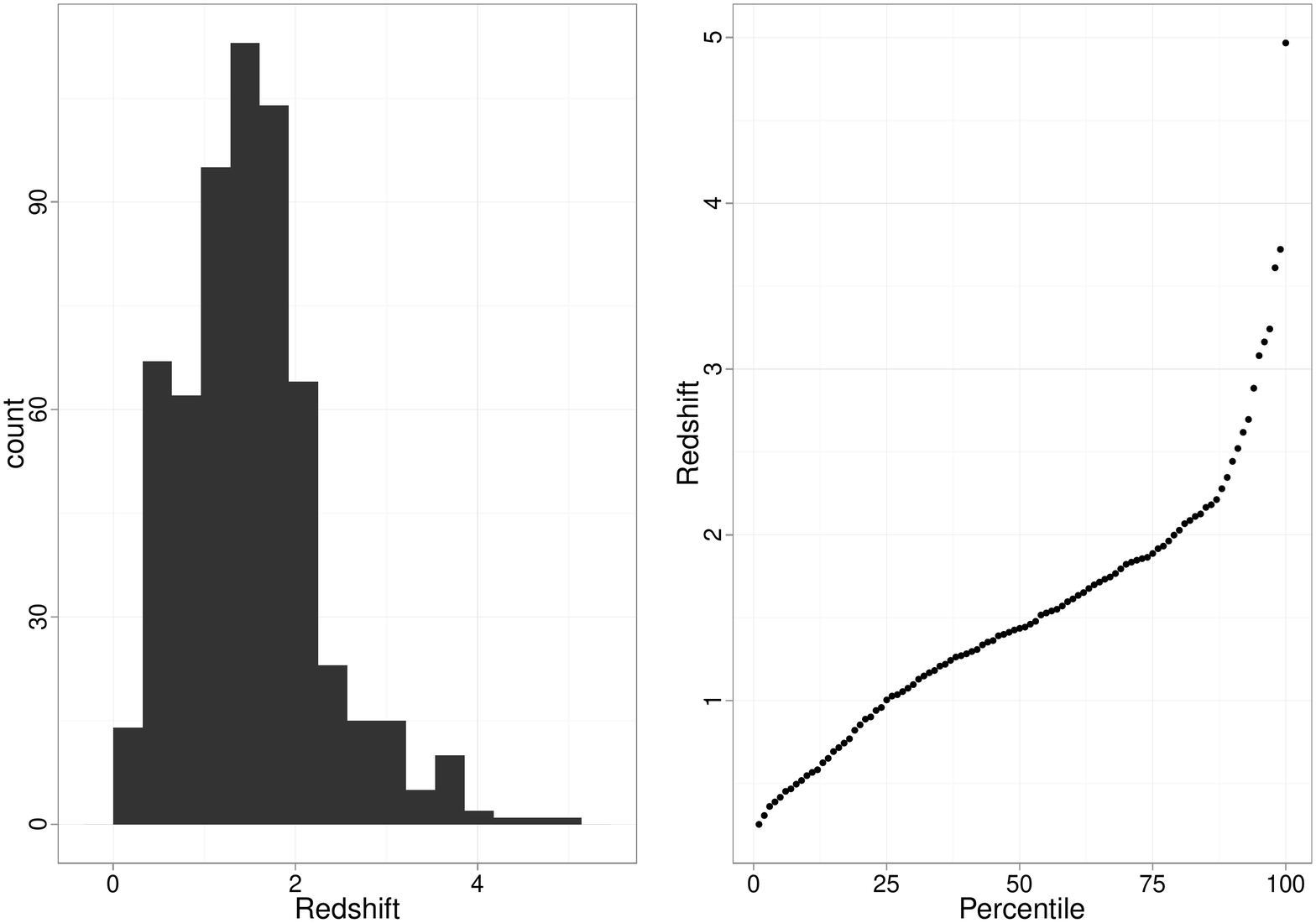} 
\caption{A histogram and a quantile plot generated using ggplot2.}
\label{fig:ggplot} 
\end{center} 
\end{figure*}

The output of the above code is shown the figure \ref{fig:base}. We now show the code needed to make a similar figure using ggplot2 library. The code is shown below and the output of the code can be seen in Figure \ref{fig:ggplot}. The code below is sufficiently commented to explain each step.

\begin{verbatim}


# Generate the same plots 
# using ggplot2 graphics
library(ggplot2)
library(gridExtra)

# Generate binwidth based on 
# Scott's formula
scott_bw <- 3.49 * sd(z_all) *
(length(z_all))^(-1/3)
z_df <- as.data.frame(z_all) 
# ggplot2 forces you to store data in a data
# frame for its functions to work, a good 
# habit in general
str(z_df) # A quick look at how the data frame looks

red_hist <- ggplot(data=z_df, aes(x=z_all))
+ geom_histogram(binwidth=scott_bw) +
xlab("Redshift") + theme_white() 
# theme_white() is a theme that creates
# publication-ready plots 

z_quant <- data.frame("Percentile"=1:100,
"Redshift"=quantile(z_df$z_all, probs=seq(1,
100, 1)/100)) # Generate quantiles to create
# a probability plot
red_qq <- ggplot(data=z_quant,
aes(x=Percentile, y=Redshift)) +
geom_point() + theme_white()

grid.arrange(red_hist, red_qq, ncol=2) 
# Display in plots in one windows split 
# into 2 columns


## Quick notes on ggplot2 layering
#
# In general, one may consider every element
# separated by a '+' to be a new layer of 
# the plot. Layers can be added or removed
# anywhere without having to alter the
# initial code block. For instance, we can
# add a diagonal line to check how good a fit is 
# uniform distribution to the data 
# by simply adding the expression 
# `geom_abline(aes(intercept=0, slope=1))' 
# to the probability plot object.
#
# Also, plot aesthetics in ggplot2 can be saved
# as a function and appended to the plot statement
# akin to `theme_white()' above. This makes it 
# convenient to create successive plots with the
# same aesthetics. The definition of `theme_white()'
# is as follows:
#
# theme_white <- function (base_size = 12, 
#                          base_family = "") {
#   theme_bw(base_size = base_size, 
#            base_family = base_family)
#      %+replace% 
#   theme(axis.title.x=element_text(size=18), 
#         axis.title.y=element_text(size=18, 
#                                   angle=90),
#         axis.text.y=element_text(angle=90),
#         axis.ticks=element_line(colour='#999999'), 
#         axis.text=element_text(size=15, 
#                                colour="black"),
#         strip.text=element_text(size=15),
#         legend.text=element_text(size=14), 
#         legend.title=element_text(size=15))
# }

\end{verbatim}

\end{document}